\documentclass[aps,prl,twocolumn,showpacs,amsmath,amssymb,floatfix,superscriptaddress]{revtex4-1}
\usepackage{graphicx}
\usepackage{dcolumn}
\usepackage{bm}
\usepackage{color}
\usepackage{epstopdf}
\usepackage{epsfig}
\usepackage[colorlinks=true, citecolor=blue, anchorcolor=green]{hyperref}

\makeatletter
\begin{document}
\preprint{APS/123-QED}

\title{Giant Enhancement of Magnonic Frequency Combs by Exceptional Points}

\author{Congyi~Wang}
\affiliation{School of Physical Science and Technology, ShanghaiTech University, Shanghai 201210, China}

\author{Jinwei~Rao}
\email{raojw@shanghaitech.edu.cn;}
\affiliation{School of Physical Science and Technology, ShanghaiTech University, Shanghai 201210, China}

\author{Zhijian~Chen}
\affiliation{School of Physical Science and Technology, ShanghaiTech University, Shanghai 201210, China}

\author{Kaixin~Zhao}
\affiliation{School of Physical Science and Technology, ShanghaiTech University, Shanghai 201210, China}

\author{Liaoxin~Sun}
\affiliation{State Key Laboratory of Infrared Physics, Shanghai Institute of Technical Physics, Chinese Academy of Sciences, Shanghai 200083, China}

\author{Bimu~Yao}\email{yaobimu@mail.sitp.ac.cn;}
\affiliation{State Key Laboratory of Infrared Physics, Shanghai Institute of Technical Physics, Chinese Academy of Sciences, Shanghai 200083, China}
\affiliation{School of Physical Science and Technology, ShanghaiTech University, Shanghai 201210, China}

\author{Tao~Yu}
\affiliation{School of Physics, Huazhong University of Science and Technology, Wuhan, 430074, China}

\author{Yi-Pu Wang}
\affiliation{Interdisciplinary Center of Quantum Information, State Key Laboratory of Modern Optical Instrumentation
and Zhejiang Province Key Laboratory of Quantum Technology and Device, School of Physics, Zhejiang University,
Hangzhou 310027, China}

\author{Wei~Lu}\email{luwei@shanghaitech.edu.cn;}
\affiliation{School of Physical Science and Technology, ShanghaiTech University, Shanghai 201210, China}
\affiliation{State Key Laboratory of Infrared Physics, Shanghai Institute of Technical Physics, Chinese Academy of Sciences, Shanghai 200083, China}

\begin{abstract}

\textbf{Abstract:} With their incomparable time–frequency accuracy, frequency combs have significantly advanced precision spectroscopy, ultra-sensitive detection, and atomic clocks. Traditional methods to create photonic, phononic, and magnonic frequency combs hinge on material nonlinearities which are often weak, necessitating high power densities to surpass their initiation thresholds, which subsequently limits their applications. Here, we introduce a novel nonlinear process to efficiently generate magnonic frequency combs (MFCs) by exploiting exceptional points (EPs) in a coupled system comprising a pump-induced magnon mode and a Kittel mode. Even without any cavity, our method greatly improves the efficiency of nonlinear frequency conversion and achieves optimal MFCs at low pump power. Additionally, our novel nonlinear process enables excellent tunability of EPs using the polarization and power of the pump, simplifying MFC generation and manipulation. Our work establishes a synergistic relationship between non-Hermitian physics and MFCs, which is advantages for coherent/quantum information processing and ultra-sensitive detection.

\end{abstract}

\maketitle

\begin{figure*} [ht]
\begin{center}
\epsfig{file=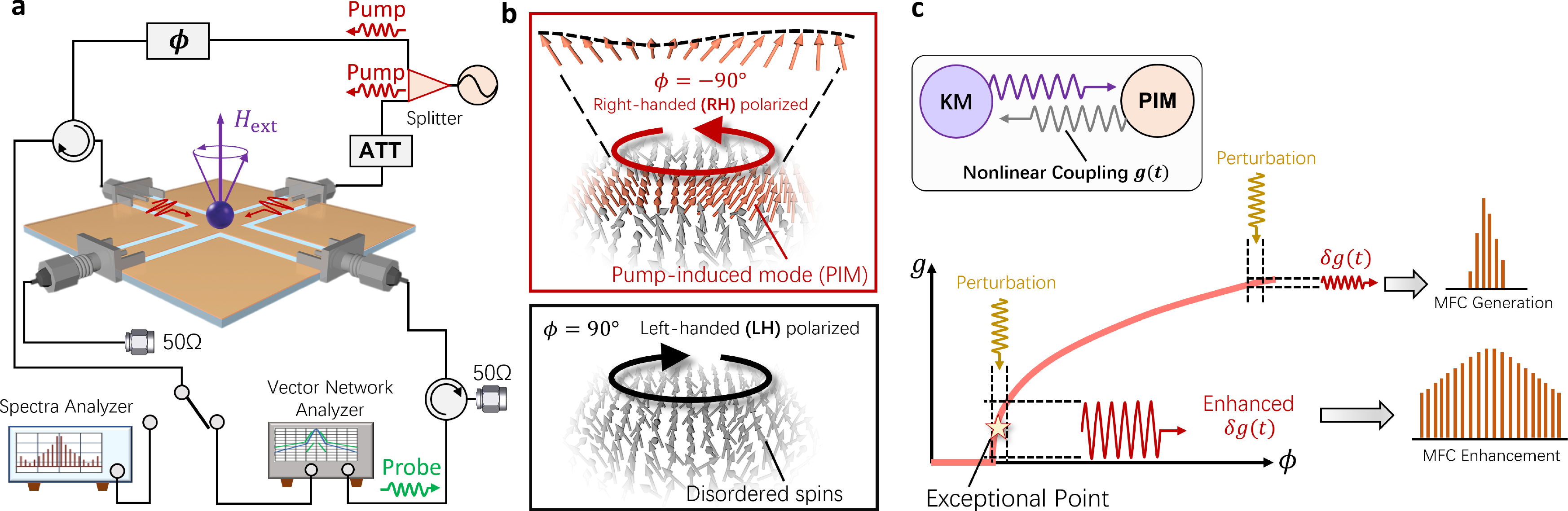,width=17.5 cm} \caption{(a) Schematic diagram of our experimental setup. The polarization of the pump field at the center of the cross is tunable by adjusting the phase difference ($\phi$) between two paths using a phase shifter. (b) Chiral control of the PIM excitation by tuning the pump field polarization. Only a pump with an RH polarization component can excite the PIM. (c) Schematic diagram of the MFC generation and enhancement. Without the probe, the PIM-KM coupling strength ($g$) is a time-independent function of the pump polarization (or $\phi$) (red solid line). With a perturbation from the probe, $g$ becomes periodic with a fluctuation $\delta g(t)$, leading to the nonlinear PIM-KM coupling. This nonlinearity reaches its maximum near EPs and hence produces the densest MFC.}\label{fig1}
\end{center}
\end{figure*}

Frequency comb refers to a spectrum of equally spaced, discrete frequency components \cite{hansch2006nobel, fortier201920, cundiff2003colloquium}. This intriguing phenomenon, which was originally discovered in optical systems \cite{udem2002optical, del2007optical, del2007optical}, has since become the foundation of a series of important technologies due to its ultrahigh time-frequency accuracy. These applications include atomic clocks \cite{ludlow2015optical}, satellite navigation \cite{lezius2016space}, molecular fingerprinting \cite{diddams2007molecular, thorpe2006broadband}, and optical spectroscopy \cite{picque2019frequency, coddington2008coherent}. Inspired by this unprecedented success, frequency combs generated by purely magnonic \cite{wang2021magnonic, hula2022spin, RaoPhysRevLett, xiong2023magnonic} and phononic \cite{cao2014phononic, ganesan2017phononic, wu2022hybridized} systems have also been achieved, recently. As the important counterparts of photonic frequency combs in microwave or radio-frequency systems, these newfound frequency combs are of significance in distance measurement\cite{doloca2010absolute}, arbitrary channel selection\cite{juan2011demonstration} and waveform generation\cite{fukushima2003optoelectronic}. The materials on which these frequency combs are implemented, however, typically have modest nonlinearities. Therefore, pump fields with high power densities are necessary for comb generations, which presents a significant challenge in the development of efficient, low-power devices.

The recently reported pump-induced magnon mode (PIM) \cite{RaoPhysRevLett, zhang2023control} and its nonlinear coupling with normal magnon modes, such as the Kittel mode (KM) in a magnetic insulator \cite{kittel1949physical, morrish2001physical}, may offer us a potential solution for overcoming these limitations. PIMs arise from the cooperative precession of unsaturated spins in a ferrimagnet when it is driven by a microwave pump; nonlinear coupling with normal magnon modes can produce a magnonic frequency comb (MFC) \cite{RaoPhysRevLett}. This nonlinear process avoids the restrictions of cavities, charge noise, and Joule heating. Its strength is sensitive to the pump power, especially at low power, making it particularly suitable for low-power MFC initiation. Furthermore, exceptional points (EPs), corresponding to spectral singularities in non-Hermitian coupled systems, have been found to exhibit stunning sensitivity to small perturbations \cite{dembowski2001experimental, miri2019exceptional, ergoktas2022topological, kononchuk2022exceptional, lai2019observation, wiersig2014enhancing, hodaei2017enhanced, kononchuk2022exceptional, zhang2019experimental, peng2014parity, chen2017exceptional}, and significant enhancement of nolinearities \cite{pick2017enhanced, PhysRevLett.117.107402, PhysRevLett.114.253601, PhysRevLett.130.110402}. We hence expect that using EPs to amplify the nonlinearity associated with a PIM may improve the MFC generation. However, leveraging EPs in a hybrid magnonic system requires intricate design and precise control. This is an area that remains unexplored and holds immense potential for following research.

In this work, we construct the EPs in a coupled PIM-KM system and achieve remarkable enhancement of MFCs. Under a two-tone drive, the PIM-KM coupling strength is periodic with time, showing resemblance to a Floquet process. As the system is tuned to EPs using the PIM's chirality, the nonlinear PIM-KM coupling is boosted, and it results in the giant enhancement of MFCs. We further use the pump power as another dimensional parameter and extend the EPs to exceptional lines in the higher-dimensional parameter space. These exceptional lines indicate the conditions required for achieving dense MFCs, providing a unique route to optimize the MFCs. Our work paves a new pathway for MFC generation, producing MFCs with a record number of teeth ($>32$) and broadening the MFC's frequency band. Our experiment combines the merits from the nonlinear coupling and peculiar EPs, which sets it apart from conventional comb generation methods based on the material nonlinearities. Although our method is demonstrated in the microwave range, it has the potential to be applied in other frequency ranges, e.g., acoustics and optics, by replicating the nonlinear coupling process. Given this, our work can advance the sensitive detection and information processing based on different combs.

We start our study by constructing EPs in a coupled PIM-KM system [Fig. \ref{fig1}(a)]. In our experimental system, a 1-mm-diameter yttrium iron garnet (YIG) sphere is placed at the center of a cross coplanar waveguide (CPW), of which each arm supports a quasi-transverse electromagnetic mode \cite{SM}. A continuous microwave pump is generated from a microwave generator (MG) and then divided into two microwave beams by a power divider. By carefully balancing their amplitudes and adjusting their phase difference ($\phi$), we can tune the polarization $\mathcal{P}$ of the pump microwaves at the center of the cross from right- to left-handedness. The PIM inherits chirality from the fixed spin-precession direction \cite{sasaki2021magnetization}, so only a pump field with a right-handed (RH) component can excite it [Fig. \ref{fig1} (b)]. This chirality is harnessed to precisely control the PIM-KM coupling \cite{lodahl2017chiral, yu2023chirality, yu2020magnon}. A weak probe with power $P_r=-18$~dBm is generated and collected by a vector network analyzer to enable transmission measurements. The radiation spectrum of the YIG sphere is measured by a signal analyzer, from which we could monitor the variations of the MFC with tuning of $\mathcal{P}$.

\begin{figure*} [ht]
\begin{center}
\epsfig{file=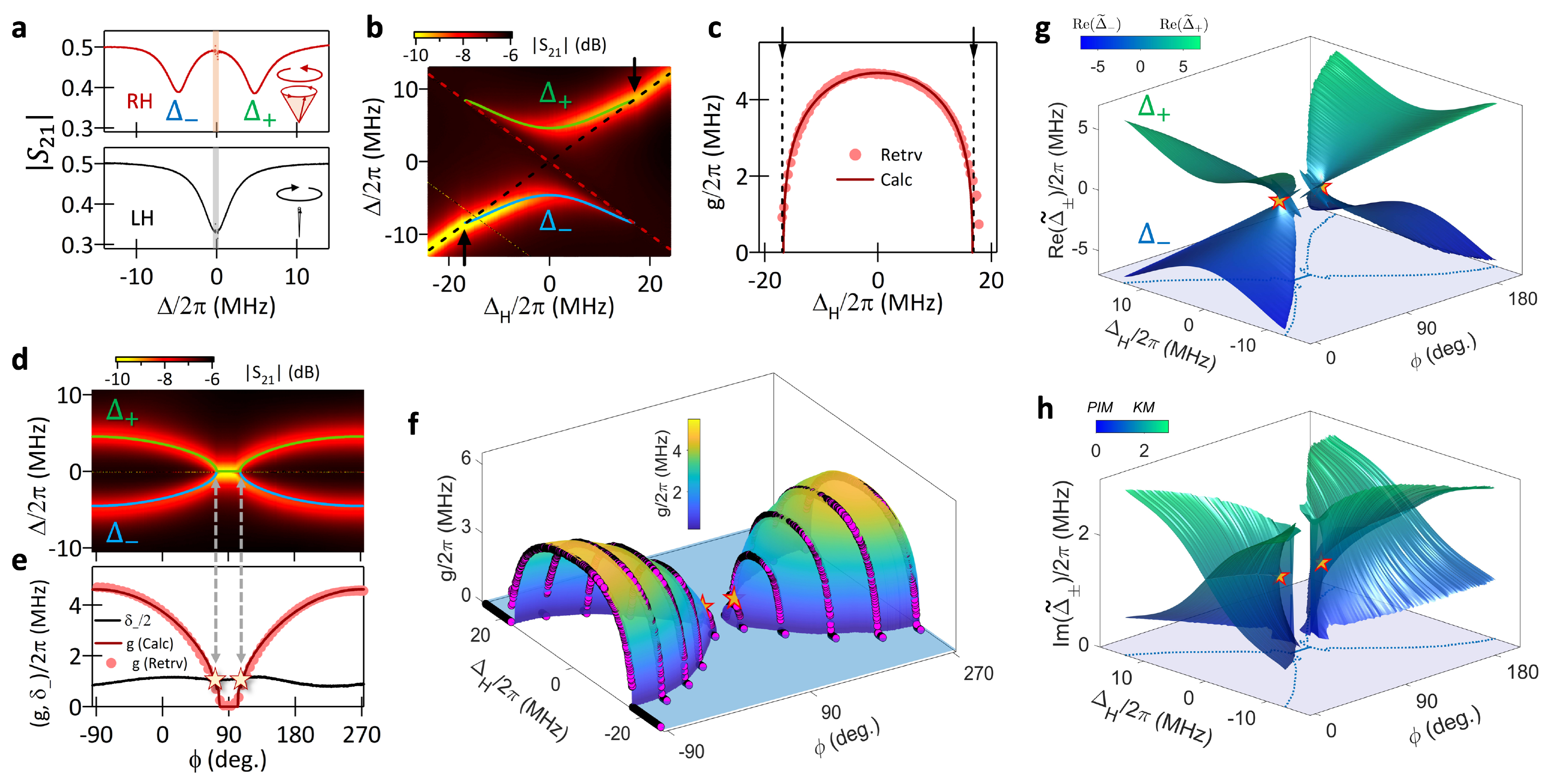,width=17.5 cm} \caption{(a) Comparison of two transmission spectra ($|S_{21}|$) respectively measured at RH and LH polarizations. (b) $|S_{21}|$ measured at different field detuning values ($\Delta_\mathrm{H}$) with a RH pump field. $\Delta=\omega_r-\omega_{rf}$ represents the probe frequency in the rotating frame. The black and red dashed lines respectively indicate the uncoupled PIM and Kittel mode. (c) $g$ at different $\Delta_\mathrm{H}$ retrieved from (b). (d) $|S_{21}|$ measured at different polarizations by setting $\Delta_\mathrm{H}=0$. The green and blue solid lines in (b) and (d) are eigenfrequencies calculated using Eq. (\ref{EGV}). (e) $g$ as a function of $\phi$. The red solid lines in (c) and (e) are calculated using Eq. (\ref{CS}). The black solid line in (e) is the half damping difference ($\delta_-/2$), whose intersections with $g$ indicate EPs (stars). (f) Evolution of $g$ plotted on a two-parameter space defined by $\Delta_\mathrm{H}$ and $\phi$. The colored surface is calculated from Eq. (\ref{CS}) using measured parameters in (e). The purple circles are experimental data. (g),(h) Riemann surfaces obtained by calculating the complex eigenfrequencies of our system. The blue dashed lines in the bottom plane indicate the diminishing boundaries of two Riemann surfaces. Stars mark two EPs on the Riemann surfaces. }\label{fig2}
\end{center}
\end{figure*}

The coupled PIM-KM system is modeled by a Hamiltonian \cite{SM}, which in a rotational frame with respect to the reference frequency $\omega_{rf}=(\omega_k+\omega_u)/2$ can be written as
\begin{equation}
    \mathcal{H}_r=\left[\begin{array}{cc}
\Delta_\mathrm{H}/2-i\alpha & g_0|\langle\hat{b}\rangle| \\
g_0|\langle\hat{b}\rangle| & -\Delta_\mathrm{H}/2-i\beta \\
\end{array}\right].
\label{HM}
\end{equation}
Here $\hat{a}$ ($\hat{a}^\dagger$) and $\hat{b}$ ($\hat{b}^\dagger$) are the annihilation (creation) operators of the KM and PIM, respectively. $\omega_k$ is the KM frequency, which is controlled by a magnetic field $\mathbf{H}_{ext}$. $\omega_u$ and $\Delta_\mathrm{H}=\omega_k-\omega_u$ represent, respectively, the pump frequency and the field detuning; and $\alpha$ and $\beta$ are the damping rates of the two modes. Since the effective spin number ($N_p$) of PIM is equal to the total number of magnons excited by the pump, i.e., $N_p=\langle\hat{b}^\dagger\hat{b}\rangle$, the PIM-KM coupling strength $g=g_0|\langle\hat{b}\rangle|\propto \sqrt{N_p}$ is tunable by the pump strength, where $g_0$ is the single spin coupling strength. When the pump is significantly stronger than the probe or the probe is not in use, $g$ is approximately constant. The coupled PIM-KM system can be linearized, so that $g$ is derived as \cite{SM}
\begin{flalign}
&g(\Delta_\mathrm{H}, \phi, A_{in}) \label{CS} & \nonumber\\
&=\sqrt{\sqrt{g_0^2\kappa_k A_u^2-(\beta-\eta A_u)^2\Delta_\mathrm{H}^2}-\alpha(\beta-\eta A_u)},
\end{flalign} 
where $A_u=A_{in}[\sqrt{(1-\sin\phi)/2}+\varepsilon]$, in which $\varepsilon$ is a small unbalanced ratio between two input pumps; $\eta$ is a fitting parameter \cite{SM}. $\kappa_k$ is the external damping of the KM arising from its coupling with the cross CPW; $A_{in}$ and $A_u$ represent, respectively, the amplitudes of the input pump and the normalized pump at the center of the cross; and $g$ is tunable by three independent parameters, $\Delta_\mathrm{H}$, $\phi$ and $A_{in}$. Such multidimensional tunability facilitates the realization of EPs and exceptional lines. 

The two eigenfrequencies of the Hamiltonian (\ref{HM}) are
\begin{equation}
    \widetilde{\Delta}_\pm=-i\frac{\delta_+}{2}\pm\sqrt{(\frac{\Delta_\mathrm{H}}{2}-i\frac{\delta_-}{2})^2+g^2},
    \label{EGV}
\end{equation}  
where $\delta_\pm=\alpha\pm\beta$ are, respectively, the damping sum and difference of the two modes. When $g$ is tuned to satisfy the critical condition with $g=\delta_-/2$ \cite{zhang2019experimental}, the two eigenfrequencies coalesce at $\Delta_\mathrm{H}=0$ and thus produce EPs.

At the EPs, the nonlinearity of the system is enhanced. Also, the small perturbation from the weak probe that is ignorable in the transmission measurements can thus no longer be neglected in the MFC measurements. The fluctuations of the PIM induced by this perturbation modify the conventionally constant coupling strength $g$ to a periodic oscillation \cite{SM}. This produces a unique nonlinearity in the PIM-KM coupling term to generate frequency conversions and thereby combs. In particular, this nonlinearity reaches its maximum at the EPs, resulting in a remarkable enhancement of the MFC [(Fig. \ref{fig1}(c)]. Next, we experimentally examine this strategy.

Figure \ref{fig2}(a) shows two transmission spectra ($|S_{21}|$) measured at RH or LH polarization, respectively. The pump frequency is 3~GHz, which matches the KM. The input pump power is $P_{in}=-4.9$~dBm. When the polarization $\mathcal{P}$ is RH, two resonance dips arising from the strong PIM-KM coupling occur in the transmission spectrum ($|S_{21}|$), indicating the two eigenfrequencies $\Delta_{\pm}$. In contrast, with the LH polarization, the PIM vanishes, leaving only a single resonance of the KM. This result demonstrates the chiral tunability of the coupled PIM-KM system.

\begin{figure} [ht]
\begin{center}
\epsfig{file=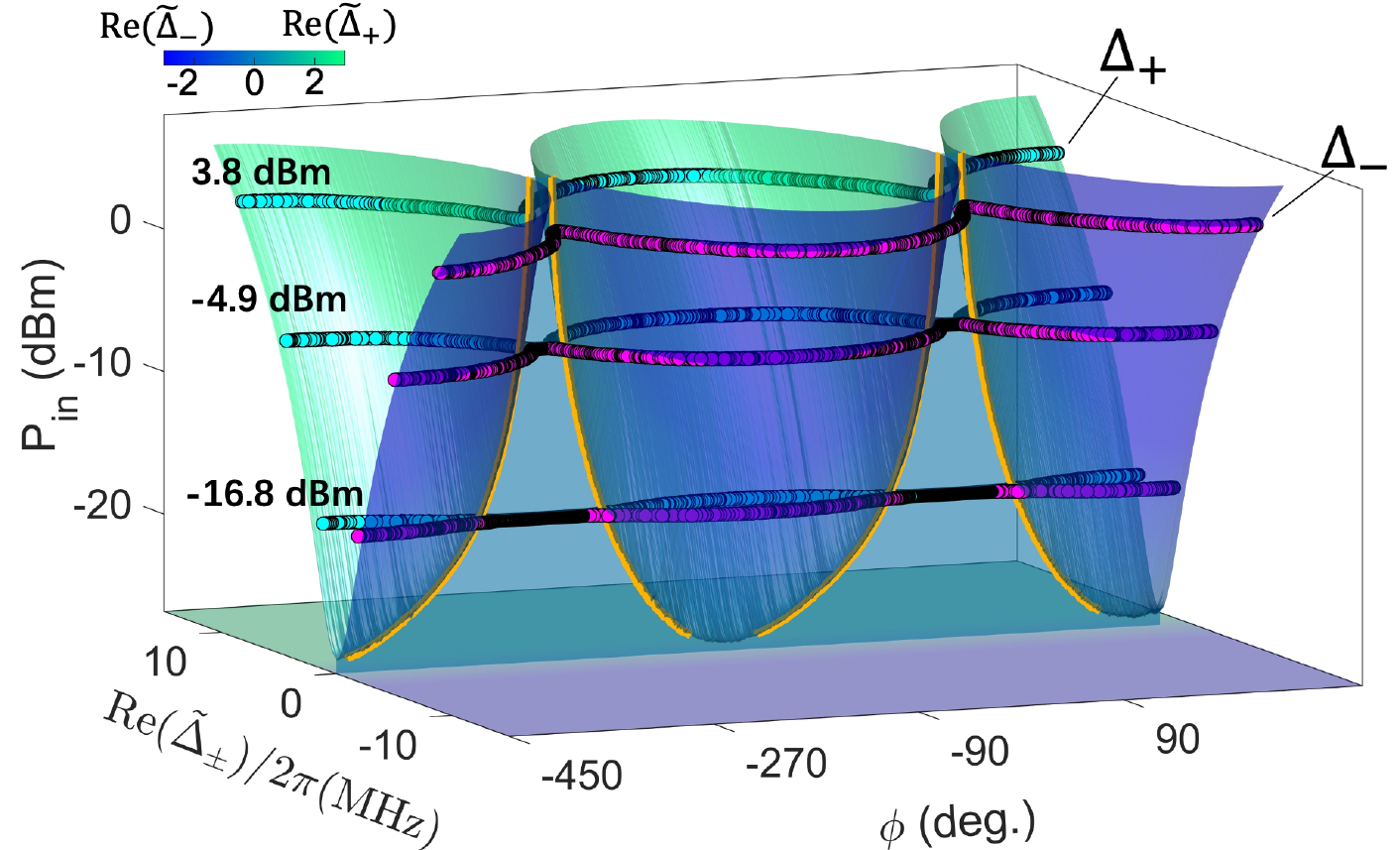,width=8.5 cm} \caption{Riemann surface obtained by calculating the real parts of the two eigenfrequencies in a parameter space defined by $P_{in}$ and $\phi$. The circles represent the resonant dips extracted from $|S_{21}|$ measured at $P_{in}=3.8$, -4.9, and -16.8 dBm. The yellow curves are exceptional lines in the parameter space.}\label{fig3}
\end{center}
\end{figure}

We next outline the PIM-KM coupling in a parameter space defined by $\Delta_\mathrm{H}$ and $\phi$ by performing two complementary measurements: 1) tuning $\Delta_\mathrm{H}$ with an RH pump, and 2) tuning $\phi$ at $\Delta_\mathrm{H}=0$. The results are summarized in Figs.~\ref{fig2}(b) and 2(c), and Figs.~\ref{fig2}(d) and 2(e). When two modes become close, as addressed by the dashed lines in Fig. \ref{fig2}(b), an anticrossing occurs because of the strong PIM-KM coupling. Two symmetric discontinuities marked by arrows imply the extinction of the PIM. The coupling strengths $g$ extracted at different $\Delta_\mathrm{H}$ values are plotted in Fig. \ref{fig2}(c). Away from the zero detuning, $g$ decreases rapidly to zero, as well captured by the calculation with Eq.~(\ref{CS}). The two zeros at $\Delta_\mathrm{H}/2\pi=\pm17$~MHz correspond exactly to the discontinuities in Fig.~\ref{fig2}(b). Next, when we tune $\mathcal{P}$ from RH to LH by changing $\phi$, two resonant dips [Fig. \ref{fig2}(d)] coalesce from the splitting. The corresponding $g$ values are plotted in Fig.~\ref{fig2}(e), which is also well reproduced by Eq.~(\ref{CS}). The intersections between $g$ and the half damping difference $\delta_-/2$, marked with stars, are the positions of the EP pair, emerging at $\phi=70^\circ$ and $112^\circ$, respectively.

\begin{figure*} [ht]
\begin{center}
\epsfig{file=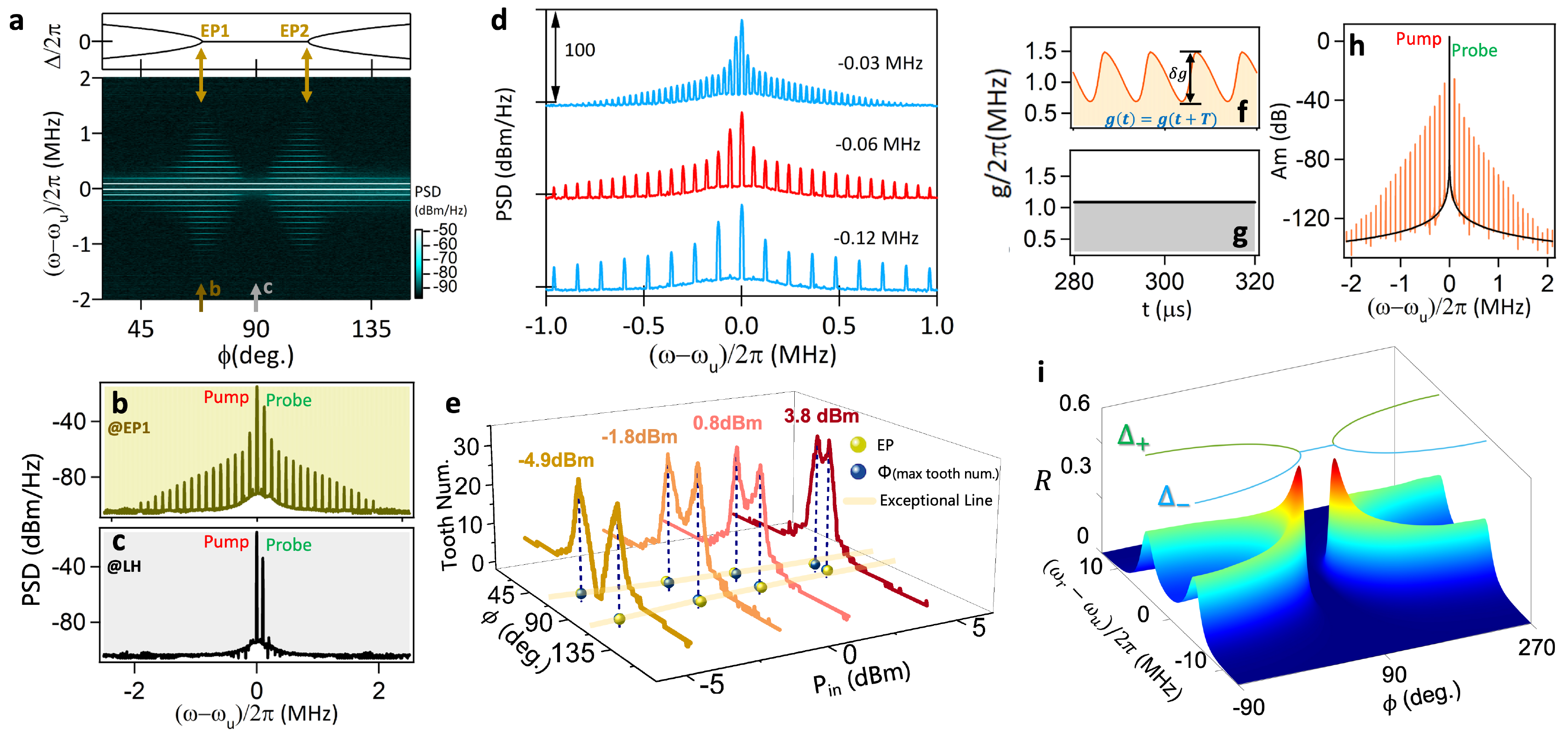,width=17.5 cm} \caption{(a) Top: evolution of the two eigenfrequencies ($\Delta_\pm$) with $\phi$. Bottom: radiation spectra measured at different $\phi$. Enhancement of the MFC occurs near two EPs. (b), (c) Two radiation spectra measured at the EP and the LH polarization, respectively. (d) Evolution of the MFC with pump-probe detuning $(\omega_r-\omega_u)/2\pi$. (e) Tooth numbers of MFCs evolving with $\phi$ respectively measured at $P_{in}=-4.9$, $-1.8$, 0.8 and 3.8 dBm. The blue balls indicate the projections of the maximal tooth numbers, which are located mear the EPs (yellow balls). Their differences are within $3^\circ$. Light orange curves represent the exceptional lines. (f), (g) Calculated $g$ at an EP with/without the probe. $\delta g$ in (f) represents the fluctuations of $g$. (h) Calculated MFC at an EP with (orange) or without (black) the probe. (i) Numerical simulated $R$, which is a function of the pump-probe detuning and $\phi$.}\label{fig4}
\end{center}
\end{figure*}

Based on the above tunability, $g$ is depicted in three-dimensional parameter space in Fig.~\ref{fig2}(f). As the system approaches the EPs (labeled by the two stars), $g$ rapidly diminishes along both $\Delta_\mathrm{H}$ and $\phi$, and hence forms a semi-cone. This feature is supported by our experimental results, for which we systematically measured the coupling constants,  as shown by the purple circles in Fig.~\ref{fig2}(f). Using Eq.~(\ref{EGV}) and the parameters in Fig.~\ref{fig2}(f), we construct the Riemann surfaces of our system by calculating its complex eigenfrequencies, as given by Figs.~\ref{fig2}(g) and (h) for the real and imaginary components, respectively. Similar to the normal EP cases, the real parts of $\widetilde{\Delta}_\pm$ repel each other and coalesce at EPs, while their imaginary parts cross. However, in contrast to  the normal EP cases, the Riemann surfaces of our system are finite and have two symmetric boundaries in the axis of $\Delta_\mathrm{H}$, as indicated by the blue dashed lines in Figs.~\ref{fig2}(g) and (h). 

In addition to the polarization, the PIM-KM coupling strength is also controllable by the pump power $P_{in}$, such that we can extend the EPs into the exceptional lines in a synthetic parameter space defined by $\phi$ and $P_{in}$. Circles in Fig.~\ref{fig3} addresses the results of three measurements performed at $P_{in}=3.8$, -4.9, and -16.8 dBm, respectively, with $\Delta_\mathrm{H}=0$. These circles represent the resonant dips extracted from $|S_{21}|$. They can reflect the system's eigenfrequencies and periodically coalesce when varying $\phi$. The calculated real parts of $\widetilde{\Delta}_\pm$ are plotted as colored surfaces in Fig.~\ref{fig3}. They coalesce at the yellow curves, which are collections of EPs that compose the exceptional lines \cite{SM}. These exceptional lines demonstrate the great convenience of constructing and manipulating the EPs in a coupled PIM-KM system, which is important for optimizing the MFCs as addressed below. 

Figure~\ref{fig4}(a) shows radiation spectra measured at different $\mathcal{P}$. The probe frequency is $\omega_r/2\pi=3.0001$ GHz, i.e., 0.1 MHz higher than the pump frequency. As the system approaches the EPs, the number of teeth in the MFC dramatically increases from several to 32. The frequency spectrum at the left EP is plotted in Fig.~\ref{fig4}(b). All teeth are symmetrically distributed relative to the pump frequency with a fixed frequency interval of 0.1 MHz, i.e., the the pump-probe detuning $(\omega_r-\omega_u)/2\pi$. When $\mathcal{P}$ is LH, the PIM vanishes, leading to the disappearance of the MFC as demonstrated in Fig. \ref{fig4}(c). In addition, the MFC can be flexibly tuned by pump-probe detuning. As it decreases, the number of teeth dramatically increases to more than 32 and the comb frequency thus becomes denser [Fig. \ref{fig4}(d)]. We repeat the measurements at different $P_{in}$ values, as shown in Fig.~\ref{fig4}(a). The variation of the number of teeth with $\phi$ is extracted and plotted as curves in Fig.~\ref{fig4}(e). The $\phi$ values corresponding to the maximal tooth numbers in all measurements are located near the exceptional lines. This means that the exceptional lines actually indicate the conditions required for achieving the maximal enhancement of MFCs at different pump powers.

The calculated $g$ values with and without a probe near an EP are plotted in Figs.~\ref{fig4}(f) and (g). Without the probe, the system steadily oscillates at the frequency of $\omega_u$, so that $g$ is constant. The system's spectrum shows a single peak, as indicated by the black curve in Fig. \ref{fig4}(h). However, under the two-tone drive, both the PIM and KM oscillate in a superposition of a steady oscillation at $\omega_u$ and fluctuations induced by the probe. Consequently, both $g(t)=g_0|\langle\hat{b}\rangle|$ and $\mathcal{H}_r(t)$ oscillate with a period of $T=2\pi/|\omega_r-\omega_u|$. The PIM-KM coupling becomes nonlinear and hence produces a MFC with an interval of $\omega_r-\omega_u$, as plotted by the orange line in Fig. \ref{fig4}(h). The periodic Hamiltonian $\mathcal{H}_r(t)$ governs the MFC generation with Floquet physics, indicating that the MFC teeth have the same physical basis as Floquet states. 

Based on the oscillating $g$, we derive the MFC expression \cite{SM}, from which we define a ratio $R$ to indicate the weight of the frequency conversion term in the MFC expression. It has a form
\begin{eqnarray}
R=\frac{g_0\overline{g}^2\sqrt{\kappa_k}A_r}{|(\omega_r-\widetilde{\omega}_+)(\omega_r-\widetilde{\omega}_-)|^2}, \label{NR}
\end{eqnarray}
where $\widetilde{\omega}_\pm$ are the system's eigenfrequencies in the lab frame, and $\overline{g}$ is the mean value of $g$. As the fluctuation of $g$ increases, $R$ becomes more prominent. Figure \ref{fig4} (i) shows the numerical simulation. At each $\phi$, $R$ has two maxima at $\omega_r=\omega_\pm$, where the MFC teeth get their maximal intensities \cite{RaoPhysRevLett, SM}. As we change $\phi$, the two maxima coalesce near EPs. $R$ attains its maximum in the whole parameter space where the MFC is greatly enhanced \cite{SM}. This result is well consistent with our measurements as shown in Figs. \ref{fig4}~(a) and (e). 

\textsl{Conclusions}.--- In summary, we have demonstrated the enhancement of MFCs by constructing EPs in a coupled PIM-KM system. Our method eliminates the restrictions of using cavities and applies an EP-enhanced nonlinear coupling process to generate combs. It overcomes the naturally weak material nonlinearity in traditional approaches, in which the generation of frequency combs demands a high power density. These properties make our method ideal for optimizing MFCs for signal processing and ultra-sensitive detection. Additionally, the EPs in our system exhibit excellent tunability and can be extended to exceptional lines in parameter space. These exceptional lines indicate all possible conditions for achieving denser MFCs via manipulating the pump. This hence provides an important guide for follow-up research in areas that may require dense MFCs. Finally, the nonlinear PIM-KM coupling renders the system's Hamiltonian periodic, leading to a connection between the MFC generation and Floquet physics. This connection opens up possibilities for breaking time-translation symmetry in nonequilibrium systems. and it has the potential to advance the exploration of new phenomena in the dynamics and stability of time crystals \cite{else2016floquet}.

\subsection*{Acknowledge}
This work has been funded by National Natural Science Foundation of China under Grants Nos.12122413, 11974369, 11991063, 12204306, 0214012051, 92265202 and 12174329, STCSM Nos.21JC1406200 and 22JC1403300, the Youth Innovation Promotion Association No. 2020247 and Strategic priority research No. XDB43010200 of CAS, the National Key R\&D Program of China (Nos. 2022YFA1404603, 2022YFA1604400, 2022YFA1405200), the SITP Independent Foundation, the Shanghai Pujiang Program (No. 22PJ1410700) and the startup grant of Huazhong University of Science and Technology (Grants Nos.3004012185 and 3004012198).

\subsection*{Author Contributions}
J.W.R. and B.M.Y. conceived this study and designed the experimental setup. C.Y.W., J.W.R. and B.M.Y. performed the measurements and data analysis. J.W.R. built the theoretical model and wrote the supplemental document. J.W.R., B.M.Y., C.Y.W., T. Y., Y.P.W., L.X.S., Z.J.C, K.X.Z and L.W. together contributed to the writing of the paper. L.W. supervised this work.


\begin{thebibliography}{44}%
\makeatletter
\providecommand \@ifxundefined [1]{%
 \@ifx{#1\undefined}
}%
\providecommand \@ifnum [1]{%
 \ifnum #1\expandafter \@firstoftwo
 \else \expandafter \@secondoftwo
 \fi
}%
\providecommand \@ifx [1]{%
 \ifx #1\expandafter \@firstoftwo
 \else \expandafter \@secondoftwo
 \fi
}%
\providecommand \natexlab [1]{#1}%
\providecommand \enquote  [1]{``#1''}%
\providecommand \bibnamefont  [1]{#1}%
\providecommand \bibfnamefont [1]{#1}%
\providecommand \citenamefont [1]{#1}%
\providecommand \href@noop [0]{\@secondoftwo}%
\providecommand \href [0]{\begingroup \@sanitize@url \@href}%
\providecommand \@href[1]{\@@startlink{#1}\@@href}%
\providecommand \@@href[1]{\endgroup#1\@@endlink}%
\providecommand \@sanitize@url [0]{\catcode `\\12\catcode `\$12\catcode
  `\&12\catcode `\#12\catcode `\^12\catcode `\_12\catcode `\%12\relax}%
\providecommand \@@startlink[1]{}%
\providecommand \@@endlink[0]{}%
\providecommand \url  [0]{\begingroup\@sanitize@url \@url }%
\providecommand \@url [1]{\endgroup\@href {#1}{\urlprefix }}%
\providecommand \urlprefix  [0]{URL }%
\providecommand \Eprint [0]{\href }%
\providecommand \doibase [0]{http://dx.doi.org/}%
\providecommand \selectlanguage [0]{\@gobble}%
\providecommand \bibinfo  [0]{\@secondoftwo}%
\providecommand \bibfield  [0]{\@secondoftwo}%
\providecommand \translation [1]{[#1]}%
\providecommand \BibitemOpen [0]{}%
\providecommand \bibitemStop [0]{}%
\providecommand \bibitemNoStop [0]{.\EOS\space}%
\providecommand \EOS [0]{\spacefactor3000\relax}%
\providecommand \BibitemShut  [1]{\csname bibitem#1\endcsname}%
\let\auto@bib@innerbib\@empty
\bibitem [{\citenamefont {H{\"a}nsch}(2006)}]{hansch2006nobel}%
  \BibitemOpen
  \bibfield  {author} {\bibinfo {author} {\bibfnamefont {T.~W.}\ \bibnamefont
  {H{\"a}nsch}},\ }\bibfield  {title} {Nobel lecture: passion for precision,\
  }\href@noop {} {\bibfield  {journal} {\bibinfo  {journal} {Reviews of Modern
  Physics}\ }\textbf {\bibinfo {volume} {78}},\ \bibinfo {pages} {1297}
  (\bibinfo {year} {2006})}\BibitemShut {NoStop}%
\bibitem [{\citenamefont {Fortier}\ and\ \citenamefont
  {Baumann}(2019)}]{fortier201920}%
  \BibitemOpen
  \bibfield  {author} {\bibinfo {author} {\bibfnamefont {T.}~\bibnamefont
  {Fortier}}\ and\ \bibinfo {author} {\bibfnamefont {E.}~\bibnamefont
  {Baumann}},\ }\bibfield  {title} {20 years of developments in optical
  frequency comb technology and applications,\ }\href@noop {} {\bibfield
  {journal} {\bibinfo  {journal} {Communications Physics}\ }\textbf {\bibinfo
  {volume} {2}},\ \bibinfo {pages} {153} (\bibinfo {year} {2019})}\BibitemShut
  {NoStop}%
\bibitem [{\citenamefont {Cundiff}\ and\ \citenamefont
  {Ye}(2003)}]{cundiff2003colloquium}%
  \BibitemOpen
  \bibfield  {author} {\bibinfo {author} {\bibfnamefont {S.~T.}\ \bibnamefont
  {Cundiff}}\ and\ \bibinfo {author} {\bibfnamefont {J.}~\bibnamefont {Ye}},\
  }\bibfield  {title} {Colloquium: Femtosecond optical frequency combs,\
  }\href@noop {} {\bibfield  {journal} {\bibinfo  {journal} {Reviews of Modern
  Physics}\ }\textbf {\bibinfo {volume} {75}},\ \bibinfo {pages} {325}
  (\bibinfo {year} {2003})}\BibitemShut {NoStop}%
\bibitem [{\citenamefont {Udem}\ \emph {et~al.}(2002)\citenamefont {Udem},
  \citenamefont {Holzwarth},\ and\ \citenamefont
  {H{\"a}nsch}}]{udem2002optical}%
  \BibitemOpen
  \bibfield  {author} {\bibinfo {author} {\bibfnamefont {T.}~\bibnamefont
  {Udem}}, \bibinfo {author} {\bibfnamefont {R.}~\bibnamefont {Holzwarth}}, \
  and\ \bibinfo {author} {\bibfnamefont {T.~W.}\ \bibnamefont {H{\"a}nsch}},\
  }\bibfield  {title} {Optical frequency metrology,\ }\href@noop {} {\bibfield
  {journal} {\bibinfo  {journal} {Nature}\ }\textbf {\bibinfo {volume} {416}},\
  \bibinfo {pages} {233} (\bibinfo {year} {2002})}\BibitemShut {NoStop}%
\bibitem [{\citenamefont {Del'Haye}\ \emph {et~al.}(2007)\citenamefont
  {Del'Haye}, \citenamefont {Schliesser}, \citenamefont {Arcizet},
  \citenamefont {Wilken}, \citenamefont {Holzwarth},\ and\ \citenamefont
  {Kippenberg}}]{del2007optical}%
  \BibitemOpen
  \bibfield  {author} {\bibinfo {author} {\bibfnamefont {P.}~\bibnamefont
  {Del'Haye}}, \bibinfo {author} {\bibfnamefont {A.}~\bibnamefont
  {Schliesser}}, \bibinfo {author} {\bibfnamefont {O.}~\bibnamefont {Arcizet}},
  \bibinfo {author} {\bibfnamefont {T.}~\bibnamefont {Wilken}}, \bibinfo
  {author} {\bibfnamefont {R.}~\bibnamefont {Holzwarth}}, \ and\ \bibinfo
  {author} {\bibfnamefont {T.~J.}\ \bibnamefont {Kippenberg}},\ }\bibfield
  {title} {Optical frequency comb generation from a monolithic microresonator,\
  }\href@noop {} {\bibfield  {journal} {\bibinfo  {journal} {Nature}\ }\textbf
  {\bibinfo {volume} {450}},\ \bibinfo {pages} {1214} (\bibinfo {year}
  {2007})}\BibitemShut {NoStop}%
\bibitem [{\citenamefont {Ludlow}\ \emph {et~al.}(2015)\citenamefont {Ludlow},
  \citenamefont {Boyd}, \citenamefont {Ye}, \citenamefont {Peik},\ and\
  \citenamefont {Schmidt}}]{ludlow2015optical}%
  \BibitemOpen
  \bibfield  {author} {\bibinfo {author} {\bibfnamefont {A.~D.}\ \bibnamefont
  {Ludlow}}, \bibinfo {author} {\bibfnamefont {M.~M.}\ \bibnamefont {Boyd}},
  \bibinfo {author} {\bibfnamefont {J.}~\bibnamefont {Ye}}, \bibinfo {author}
  {\bibfnamefont {E.}~\bibnamefont {Peik}}, \ and\ \bibinfo {author}
  {\bibfnamefont {P.~O.}\ \bibnamefont {Schmidt}},\ }\bibfield  {title}
  {Optical atomic clocks,\ }\href@noop {} {\bibfield  {journal} {\bibinfo
  {journal} {Reviews of Modern Physics}\ }\textbf {\bibinfo {volume} {87}},\
  \bibinfo {pages} {637} (\bibinfo {year} {2015})}\BibitemShut {NoStop}%
\bibitem [{\citenamefont {Lezius}\ \emph {et~al.}(2016)\citenamefont {Lezius},
  \citenamefont {Wilken}, \citenamefont {Deutsch}, \citenamefont {Giunta},
  \citenamefont {Mandel}, \citenamefont {Thaller}, \citenamefont {Schkolnik},
  \citenamefont {Schiemangk}, \citenamefont {Dinkelaker}, \citenamefont
  {Kohfeldt} \emph {et~al.}}]{lezius2016space}%
  \BibitemOpen
  \bibfield  {author} {\bibinfo {author} {\bibfnamefont {M.}~\bibnamefont
  {Lezius}}, \bibinfo {author} {\bibfnamefont {T.}~\bibnamefont {Wilken}},
  \bibinfo {author} {\bibfnamefont {C.}~\bibnamefont {Deutsch}}, \bibinfo
  {author} {\bibfnamefont {M.}~\bibnamefont {Giunta}}, \bibinfo {author}
  {\bibfnamefont {O.}~\bibnamefont {Mandel}}, \bibinfo {author} {\bibfnamefont
  {A.}~\bibnamefont {Thaller}}, \bibinfo {author} {\bibfnamefont
  {V.}~\bibnamefont {Schkolnik}}, \bibinfo {author} {\bibfnamefont
  {M.}~\bibnamefont {Schiemangk}}, \bibinfo {author} {\bibfnamefont
  {A.}~\bibnamefont {Dinkelaker}}, \bibinfo {author} {\bibfnamefont
  {A.}~\bibnamefont {Kohfeldt}},  \emph {et~al.},\ }\bibfield  {title}
  {Space-borne frequency comb metrology,\ }\href@noop {} {\bibfield  {journal}
  {\bibinfo  {journal} {Optica}\ }\textbf {\bibinfo {volume} {3}},\ \bibinfo
  {pages} {1381} (\bibinfo {year} {2016})}\BibitemShut {NoStop}%
\bibitem [{\citenamefont {Diddams}\ \emph {et~al.}(2007)\citenamefont
  {Diddams}, \citenamefont {Hollberg},\ and\ \citenamefont
  {Mbele}}]{diddams2007molecular}%
  \BibitemOpen
  \bibfield  {author} {\bibinfo {author} {\bibfnamefont {S.~A.}\ \bibnamefont
  {Diddams}}, \bibinfo {author} {\bibfnamefont {L.}~\bibnamefont {Hollberg}}, \
  and\ \bibinfo {author} {\bibfnamefont {V.}~\bibnamefont {Mbele}},\ }\bibfield
   {title} {Molecular fingerprinting with the resolved modes of a femtosecond
  laser frequency comb,\ }\href@noop {} {\bibfield  {journal} {\bibinfo
  {journal} {Nature}\ }\textbf {\bibinfo {volume} {445}},\ \bibinfo {pages}
  {627} (\bibinfo {year} {2007})}\BibitemShut {NoStop}%
\bibitem [{\citenamefont {Thorpe}\ \emph {et~al.}(2006)\citenamefont {Thorpe},
  \citenamefont {Moll}, \citenamefont {Jones}, \citenamefont {Safdi},\ and\
  \citenamefont {Ye}}]{thorpe2006broadband}%
  \BibitemOpen
  \bibfield  {author} {\bibinfo {author} {\bibfnamefont {M.~J.}\ \bibnamefont
  {Thorpe}}, \bibinfo {author} {\bibfnamefont {K.~D.}\ \bibnamefont {Moll}},
  \bibinfo {author} {\bibfnamefont {R.~J.}\ \bibnamefont {Jones}}, \bibinfo
  {author} {\bibfnamefont {B.}~\bibnamefont {Safdi}}, \ and\ \bibinfo {author}
  {\bibfnamefont {J.}~\bibnamefont {Ye}},\ }\bibfield  {title} {Broadband
  cavity ringdown spectroscopy for sensitive and rapid molecular detection,\
  }\href@noop {} {\bibfield  {journal} {\bibinfo  {journal} {Science}\ }\textbf
  {\bibinfo {volume} {311}},\ \bibinfo {pages} {1595} (\bibinfo {year}
  {2006})}\BibitemShut {NoStop}%
\bibitem [{\citenamefont {Picqu{\'e}}\ and\ \citenamefont
  {H{\"a}nsch}(2019)}]{picque2019frequency}%
  \BibitemOpen
  \bibfield  {author} {\bibinfo {author} {\bibfnamefont {N.}~\bibnamefont
  {Picqu{\'e}}}\ and\ \bibinfo {author} {\bibfnamefont {T.~W.}\ \bibnamefont
  {H{\"a}nsch}},\ }\bibfield  {title} {Frequency comb spectroscopy,\
  }\href@noop {} {\bibfield  {journal} {\bibinfo  {journal} {Nature Photonics}\
  }\textbf {\bibinfo {volume} {13}},\ \bibinfo {pages} {146} (\bibinfo {year}
  {2019})}\BibitemShut {NoStop}%
\bibitem [{\citenamefont {Coddington}\ \emph {et~al.}(2008)\citenamefont
  {Coddington}, \citenamefont {Swann},\ and\ \citenamefont
  {Newbury}}]{coddington2008coherent}%
  \BibitemOpen
  \bibfield  {author} {\bibinfo {author} {\bibfnamefont {I.}~\bibnamefont
  {Coddington}}, \bibinfo {author} {\bibfnamefont {W.~C.}\ \bibnamefont
  {Swann}}, \ and\ \bibinfo {author} {\bibfnamefont {N.~R.}\ \bibnamefont
  {Newbury}},\ }\bibfield  {title} {Coherent multiheterodyne spectroscopy using
  stabilized optical frequency combs,\ }\href@noop {} {\bibfield  {journal}
  {\bibinfo  {journal} {Phys. Rev. Lett.}\ }\textbf {\bibinfo {volume} {100}},\
  \bibinfo {pages} {013902} (\bibinfo {year} {2008})}\BibitemShut {NoStop}%
\bibitem [{\citenamefont {Wang}\ \emph {et~al.}(2021)\citenamefont {Wang},
  \citenamefont {Yuan}, \citenamefont {Cao}, \citenamefont {Li}, \citenamefont
  {Duine},\ and\ \citenamefont {Yan}}]{wang2021magnonic}%
  \BibitemOpen
  \bibfield  {author} {\bibinfo {author} {\bibfnamefont {Z.}~\bibnamefont
  {Wang}}, \bibinfo {author} {\bibfnamefont {H.}~\bibnamefont {Yuan}}, \bibinfo
  {author} {\bibfnamefont {Y.}~\bibnamefont {Cao}}, \bibinfo {author}
  {\bibfnamefont {Z.-X.}\ \bibnamefont {Li}}, \bibinfo {author} {\bibfnamefont
  {R.~A.}\ \bibnamefont {Duine}}, \ and\ \bibinfo {author} {\bibfnamefont
  {P.}~\bibnamefont {Yan}},\ }\bibfield  {title} {Magnonic frequency comb
  through nonlinear magnon-skyrmion scattering,\ }\href@noop {} {\bibfield
  {journal} {\bibinfo  {journal} {Phys. Rev. Lett.}\ }\textbf {\bibinfo
  {volume} {127}},\ \bibinfo {pages} {037202} (\bibinfo {year}
  {2021})}\BibitemShut {NoStop}%
\bibitem [{\citenamefont {Hula}\ \emph {et~al.}(2022)\citenamefont {Hula},
  \citenamefont {Schultheiss}, \citenamefont {Goncalves}, \citenamefont
  {K{\"o}rber}, \citenamefont {Bejarano}, \citenamefont {Copus}, \citenamefont
  {Flacke}, \citenamefont {Liensberger}, \citenamefont {Buzdakov},
  \citenamefont {K{\'a}kay} \emph {et~al.}}]{hula2022spin}%
  \BibitemOpen
  \bibfield  {author} {\bibinfo {author} {\bibfnamefont {T.}~\bibnamefont
  {Hula}}, \bibinfo {author} {\bibfnamefont {K.}~\bibnamefont {Schultheiss}},
  \bibinfo {author} {\bibfnamefont {F.~J.~T.}\ \bibnamefont {Goncalves}},
  \bibinfo {author} {\bibfnamefont {L.}~\bibnamefont {K{\"o}rber}}, \bibinfo
  {author} {\bibfnamefont {M.}~\bibnamefont {Bejarano}}, \bibinfo {author}
  {\bibfnamefont {M.}~\bibnamefont {Copus}}, \bibinfo {author} {\bibfnamefont
  {L.}~\bibnamefont {Flacke}}, \bibinfo {author} {\bibfnamefont
  {L.}~\bibnamefont {Liensberger}}, \bibinfo {author} {\bibfnamefont
  {A.}~\bibnamefont {Buzdakov}}, \bibinfo {author} {\bibfnamefont
  {A.}~\bibnamefont {K{\'a}kay}},  \emph {et~al.},\ }\bibfield  {title}
  {Spin-wave frequency combs,\ }\href@noop {} {\bibfield  {journal} {\bibinfo
  {journal} {Applied Physics Letters}\ }\textbf {\bibinfo {volume} {121}},\
  \bibinfo {pages} {112404} (\bibinfo {year} {2022})}\BibitemShut {NoStop}%
\bibitem [{\citenamefont {Rao}\ \emph {et~al.}(2023)\citenamefont {Rao},
  \citenamefont {Yao}, \citenamefont {Wang}, \citenamefont {Zhang},
  \citenamefont {Yu},\ and\ \citenamefont {Lu}}]{RaoPhysRevLett}%
  \BibitemOpen
  \bibfield  {author} {\bibinfo {author} {\bibfnamefont {J.~W.}\ \bibnamefont
  {Rao}}, \bibinfo {author} {\bibfnamefont {B.}~\bibnamefont {Yao}}, \bibinfo
  {author} {\bibfnamefont {C.~Y.}\ \bibnamefont {Wang}}, \bibinfo {author}
  {\bibfnamefont {C.}~\bibnamefont {Zhang}}, \bibinfo {author} {\bibfnamefont
  {T.}~\bibnamefont {Yu}}, \ and\ \bibinfo {author} {\bibfnamefont
  {W.}~\bibnamefont {Lu}},\ }\bibfield  {title} {Unveiling a pump-induced
  magnon mode via its strong interaction with walker modes,\ }\href@noop {}
  {\bibfield  {journal} {\bibinfo  {journal} {Phys. Rev. Lett.}\ }\textbf
  {\bibinfo {volume} {130}},\ \bibinfo {pages} {046705} (\bibinfo {year}
  {2023})}\BibitemShut {NoStop}%
\bibitem [{\citenamefont {Xiong}(2023)}]{xiong2023magnonic}%
  \BibitemOpen
  \bibfield  {author} {\bibinfo {author} {\bibfnamefont {H.}~\bibnamefont
  {Xiong}},\ }\bibfield  {title} {Magnonic frequency combs based on the
  resonantly enhanced magnetostrictive effect,\ }\href@noop {} {\bibfield
  {journal} {\bibinfo  {journal} {Fundamental Research}\ }\textbf {\bibinfo
  {volume} {3}},\ \bibinfo {pages} {8} (\bibinfo {year} {2023})}\BibitemShut
  {NoStop}%
\bibitem [{\citenamefont {Cao}\ \emph {et~al.}(2014)\citenamefont {Cao},
  \citenamefont {Qi}, \citenamefont {Peng}, \citenamefont {Wang},\ and\
  \citenamefont {Schmelcher}}]{cao2014phononic}%
  \BibitemOpen
  \bibfield  {author} {\bibinfo {author} {\bibfnamefont {L.}~\bibnamefont
  {Cao}}, \bibinfo {author} {\bibfnamefont {D.}~\bibnamefont {Qi}}, \bibinfo
  {author} {\bibfnamefont {R.}~\bibnamefont {Peng}}, \bibinfo {author}
  {\bibfnamefont {M.}~\bibnamefont {Wang}}, \ and\ \bibinfo {author}
  {\bibfnamefont {P.}~\bibnamefont {Schmelcher}},\ }\bibfield  {title}
  {Phononic frequency combs through nonlinear resonances,\ }\href@noop {}
  {\bibfield  {journal} {\bibinfo  {journal} {Phys. Rev. Lett.}\ }\textbf
  {\bibinfo {volume} {112}},\ \bibinfo {pages} {075505} (\bibinfo {year}
  {2014})}\BibitemShut {NoStop}%
\bibitem [{\citenamefont {Ganesan}\ \emph {et~al.}(2017)\citenamefont
  {Ganesan}, \citenamefont {Do},\ and\ \citenamefont
  {Seshia}}]{ganesan2017phononic}%
  \BibitemOpen
  \bibfield  {author} {\bibinfo {author} {\bibfnamefont {A.}~\bibnamefont
  {Ganesan}}, \bibinfo {author} {\bibfnamefont {C.}~\bibnamefont {Do}}, \ and\
  \bibinfo {author} {\bibfnamefont {A.}~\bibnamefont {Seshia}},\ }\bibfield
  {title} {Phononic frequency comb via intrinsic three-wave mixing,\
  }\href@noop {} {\bibfield  {journal} {\bibinfo  {journal} {Phys. Rev. Lett.}\
  }\textbf {\bibinfo {volume} {118}},\ \bibinfo {pages} {033903} (\bibinfo
  {year} {2017})}\BibitemShut {NoStop}%
\bibitem [{\citenamefont {Wu}\ \emph {et~al.}(2022)\citenamefont {Wu},
  \citenamefont {Liu}, \citenamefont {Liu}, \citenamefont {Wang}, \citenamefont
  {Chen},\ and\ \citenamefont {Li}}]{wu2022hybridized}%
  \BibitemOpen
  \bibfield  {author} {\bibinfo {author} {\bibfnamefont {S.}~\bibnamefont
  {Wu}}, \bibinfo {author} {\bibfnamefont {Y.}~\bibnamefont {Liu}}, \bibinfo
  {author} {\bibfnamefont {Q.}~\bibnamefont {Liu}}, \bibinfo {author}
  {\bibfnamefont {S.-P.}\ \bibnamefont {Wang}}, \bibinfo {author}
  {\bibfnamefont {Z.}~\bibnamefont {Chen}}, \ and\ \bibinfo {author}
  {\bibfnamefont {T.}~\bibnamefont {Li}},\ }\bibfield  {title} {Hybridized
  frequency combs in multimode cavity electromechanical system,\ }\href@noop {}
  {\bibfield  {journal} {\bibinfo  {journal} {Phys. Rev. Lett.}\ }\textbf
  {\bibinfo {volume} {128}},\ \bibinfo {pages} {153901} (\bibinfo {year}
  {2022})}\BibitemShut {NoStop}%
\bibitem [{\citenamefont {Doloca}\ \emph {et~al.}(2010)\citenamefont {Doloca},
  \citenamefont {Meiners-Hagen}, \citenamefont {Wedde}, \citenamefont
  {Pollinger},\ and\ \citenamefont {Abou-Zeid}}]{doloca2010absolute}%
  \BibitemOpen
  \bibfield  {author} {\bibinfo {author} {\bibfnamefont {N.~R.}\ \bibnamefont
  {Doloca}}, \bibinfo {author} {\bibfnamefont {K.}~\bibnamefont
  {Meiners-Hagen}}, \bibinfo {author} {\bibfnamefont {M.}~\bibnamefont
  {Wedde}}, \bibinfo {author} {\bibfnamefont {F.}~\bibnamefont {Pollinger}}, \
  and\ \bibinfo {author} {\bibfnamefont {A.}~\bibnamefont {Abou-Zeid}},\
  }\bibfield  {title} {Absolute distance measurement system using a femtosecond
  laser as a modulator,\ }\href@noop {} {\bibfield  {journal} {\bibinfo
  {journal} {Measurement Science and Technology}\ }\textbf {\bibinfo {volume}
  {21}},\ \bibinfo {pages} {115302} (\bibinfo {year} {2010})}\BibitemShut
  {NoStop}%
\bibitem [{\citenamefont {Juan}\ and\ \citenamefont
  {Lin}(2011)}]{juan2011demonstration}%
  \BibitemOpen
  \bibfield  {author} {\bibinfo {author} {\bibfnamefont {Y.-S.}\ \bibnamefont
  {Juan}}\ and\ \bibinfo {author} {\bibfnamefont {F.-Y.}\ \bibnamefont {Lin}},\
  }\bibfield  {title} {Demonstration of arbitrary channel selection utilizing a
  pulse-injected semiconductor laser with a phase-locked loop,\ }\href@noop {}
  {\bibfield  {journal} {\bibinfo  {journal} {Optics Express}\ }\textbf
  {\bibinfo {volume} {19}},\ \bibinfo {pages} {1057} (\bibinfo {year}
  {2011})}\BibitemShut {NoStop}%
\bibitem [{\citenamefont {Fukushima}\ \emph {et~al.}(2003)\citenamefont
  {Fukushima}, \citenamefont {Silva}, \citenamefont {Muramoto},\ and\
  \citenamefont {Seeds}}]{fukushima2003optoelectronic}%
  \BibitemOpen
  \bibfield  {author} {\bibinfo {author} {\bibfnamefont {S.}~\bibnamefont
  {Fukushima}}, \bibinfo {author} {\bibfnamefont {C.}~\bibnamefont {Silva}},
  \bibinfo {author} {\bibfnamefont {Y.}~\bibnamefont {Muramoto}}, \ and\
  \bibinfo {author} {\bibfnamefont {A.~J.}\ \bibnamefont {Seeds}},\ }\bibfield
  {title} {Optoelectronic millimeter-wave synthesis using an optical frequency
  comb generator, optically injection locked lasers, and a unitraveling-carrier
  photodiode,\ }\href@noop {} {\bibfield  {journal} {\bibinfo  {journal}
  {Journal of lightwave technology}\ }\textbf {\bibinfo {volume} {21}},\
  \bibinfo {pages} {3043} (\bibinfo {year} {2003})}\BibitemShut {NoStop}%
\bibitem [{\citenamefont {Zhang}\ \emph {et~al.}(2023)\citenamefont {Zhang},
  \citenamefont {Rao}, \citenamefont {Wang}, \citenamefont {Chen},
  \citenamefont {Zhao}, \citenamefont {Yao}, \citenamefont {Xu},\ and\
  \citenamefont {Lu}}]{zhang2023control}%
  \BibitemOpen
  \bibfield  {author} {\bibinfo {author} {\bibfnamefont {C.}~\bibnamefont
  {Zhang}}, \bibinfo {author} {\bibfnamefont {J.}~\bibnamefont {Rao}}, \bibinfo
  {author} {\bibfnamefont {C.}~\bibnamefont {Wang}}, \bibinfo {author}
  {\bibfnamefont {Z.}~\bibnamefont {Chen}}, \bibinfo {author} {\bibfnamefont
  {K.}~\bibnamefont {Zhao}}, \bibinfo {author} {\bibfnamefont {B.}~\bibnamefont
  {Yao}}, \bibinfo {author} {\bibfnamefont {X.-G.}\ \bibnamefont {Xu}}, \ and\
  \bibinfo {author} {\bibfnamefont {W.}~\bibnamefont {Lu}},\ }\bibfield
  {title} {Control of the magnon-polariton hybridization with a microwave
  pump,\ }\href@noop {} {\bibfield  {journal} {\bibinfo  {journal} {arXiv
  preprint arXiv:2302.08665}\ } (\bibinfo {year} {2023})}\BibitemShut {NoStop}%
\bibitem [{\citenamefont {Kittel}(1949)}]{kittel1949physical}%
  \BibitemOpen
  \bibfield  {author} {\bibinfo {author} {\bibfnamefont {C.}~\bibnamefont
  {Kittel}},\ }\bibfield  {title} {Physical theory of ferromagnetic domains,\
  }\href@noop {} {\bibfield  {journal} {\bibinfo  {journal} {Reviews of Modern
  Physics}\ }\textbf {\bibinfo {volume} {21}},\ \bibinfo {pages} {541}
  (\bibinfo {year} {1949})}\BibitemShut {NoStop}%
\bibitem [{\citenamefont {Morrish}(2001)}]{morrish2001physical}%
  \BibitemOpen
  \bibfield  {author} {\bibinfo {author} {\bibfnamefont {A.~H.}\ \bibnamefont
  {Morrish}},\ }\href@noop {} {\emph {\bibinfo {title} {The physical principles
  of magnetism}}}\ (\bibinfo {year} {2001})\BibitemShut {NoStop}%
\bibitem [{\citenamefont {Dembowski}\ \emph {et~al.}(2001)\citenamefont
  {Dembowski}, \citenamefont {Gr{\"a}f}, \citenamefont {Harney}, \citenamefont
  {Heine}, \citenamefont {Heiss}, \citenamefont {Rehfeld},\ and\ \citenamefont
  {Richter}}]{dembowski2001experimental}%
  \BibitemOpen
  \bibfield  {author} {\bibinfo {author} {\bibfnamefont {C.}~\bibnamefont
  {Dembowski}}, \bibinfo {author} {\bibfnamefont {H.-D.}\ \bibnamefont
  {Gr{\"a}f}}, \bibinfo {author} {\bibfnamefont {H.}~\bibnamefont {Harney}},
  \bibinfo {author} {\bibfnamefont {A.}~\bibnamefont {Heine}}, \bibinfo
  {author} {\bibfnamefont {W.}~\bibnamefont {Heiss}}, \bibinfo {author}
  {\bibfnamefont {H.}~\bibnamefont {Rehfeld}}, \ and\ \bibinfo {author}
  {\bibfnamefont {A.}~\bibnamefont {Richter}},\ }\bibfield  {title}
  {Experimental observation of the topological structure of exceptional
  points,\ }\href@noop {} {\bibfield  {journal} {\bibinfo  {journal} {Phys.
  Rev. Lett.}\ }\textbf {\bibinfo {volume} {86}},\ \bibinfo {pages} {787}
  (\bibinfo {year} {2001})}\BibitemShut {NoStop}%
\bibitem [{\citenamefont {Miri}\ and\ \citenamefont
  {Al{\`u}}(2019)}]{miri2019exceptional}%
  \BibitemOpen
  \bibfield  {author} {\bibinfo {author} {\bibfnamefont {M.-A.}\ \bibnamefont
  {Miri}}\ and\ \bibinfo {author} {\bibfnamefont {A.}~\bibnamefont {Al{\`u}}},\
  }\bibfield  {title} {Exceptional points in optics and photonics,\ }\href@noop
  {} {\bibfield  {journal} {\bibinfo  {journal} {Science}\ }\textbf {\bibinfo
  {volume} {363}},\ \bibinfo {pages} {eaar7709} (\bibinfo {year}
  {2019})}\BibitemShut {NoStop}%
\bibitem [{\citenamefont {Ergoktas}\ \emph {et~al.}(2022)\citenamefont
  {Ergoktas}, \citenamefont {Soleymani}, \citenamefont {Kakenov}, \citenamefont
  {Wang}, \citenamefont {Smith}, \citenamefont {Bakan}, \citenamefont {Balci},
  \citenamefont {Principi}, \citenamefont {Novoselov}, \citenamefont {Ozdemir}
  \emph {et~al.}}]{ergoktas2022topological}%
  \BibitemOpen
  \bibfield  {author} {\bibinfo {author} {\bibfnamefont {M.~S.}\ \bibnamefont
  {Ergoktas}}, \bibinfo {author} {\bibfnamefont {S.}~\bibnamefont {Soleymani}},
  \bibinfo {author} {\bibfnamefont {N.}~\bibnamefont {Kakenov}}, \bibinfo
  {author} {\bibfnamefont {K.}~\bibnamefont {Wang}}, \bibinfo {author}
  {\bibfnamefont {T.~B.}\ \bibnamefont {Smith}}, \bibinfo {author}
  {\bibfnamefont {G.}~\bibnamefont {Bakan}}, \bibinfo {author} {\bibfnamefont
  {S.}~\bibnamefont {Balci}}, \bibinfo {author} {\bibfnamefont
  {A.}~\bibnamefont {Principi}}, \bibinfo {author} {\bibfnamefont {K.~S.}\
  \bibnamefont {Novoselov}}, \bibinfo {author} {\bibfnamefont {S.~K.}\
  \bibnamefont {Ozdemir}},  \emph {et~al.},\ }\bibfield  {title} {Topological
  engineering of terahertz light using electrically tunable exceptional point
  singularities,\ }\href@noop {} {\bibfield  {journal} {\bibinfo  {journal}
  {Science}\ }\textbf {\bibinfo {volume} {376}},\ \bibinfo {pages} {184}
  (\bibinfo {year} {2022})}\BibitemShut {NoStop}%
\bibitem [{\citenamefont {Kononchuk}\ \emph {et~al.}(2022)\citenamefont
  {Kononchuk}, \citenamefont {Cai}, \citenamefont {Ellis}, \citenamefont
  {Thevamaran},\ and\ \citenamefont {Kottos}}]{kononchuk2022exceptional}%
  \BibitemOpen
  \bibfield  {author} {\bibinfo {author} {\bibfnamefont {R.}~\bibnamefont
  {Kononchuk}}, \bibinfo {author} {\bibfnamefont {J.}~\bibnamefont {Cai}},
  \bibinfo {author} {\bibfnamefont {F.}~\bibnamefont {Ellis}}, \bibinfo
  {author} {\bibfnamefont {R.}~\bibnamefont {Thevamaran}}, \ and\ \bibinfo
  {author} {\bibfnamefont {T.}~\bibnamefont {Kottos}},\ }\bibfield  {title}
  {Exceptional-point-based accelerometers with enhanced signal-to-noise ratio,\
  }\href@noop {} {\bibfield  {journal} {\bibinfo  {journal} {Nature}\ }\textbf
  {\bibinfo {volume} {607}},\ \bibinfo {pages} {697} (\bibinfo {year}
  {2022})}\BibitemShut {NoStop}%
\bibitem [{\citenamefont {Lai}\ \emph {et~al.}(2019)\citenamefont {Lai},
  \citenamefont {Lu}, \citenamefont {Suh}, \citenamefont {Yuan},\ and\
  \citenamefont {Vahala}}]{lai2019observation}%
  \BibitemOpen
  \bibfield  {author} {\bibinfo {author} {\bibfnamefont {Y.-H.}\ \bibnamefont
  {Lai}}, \bibinfo {author} {\bibfnamefont {Y.-K.}\ \bibnamefont {Lu}},
  \bibinfo {author} {\bibfnamefont {M.-G.}\ \bibnamefont {Suh}}, \bibinfo
  {author} {\bibfnamefont {Z.}~\bibnamefont {Yuan}}, \ and\ \bibinfo {author}
  {\bibfnamefont {K.}~\bibnamefont {Vahala}},\ }\bibfield  {title} {Observation
  of the exceptional-point-enhanced sagnac effect,\ }\href@noop {} {\bibfield
  {journal} {\bibinfo  {journal} {Nature}\ }\textbf {\bibinfo {volume} {576}},\
  \bibinfo {pages} {65} (\bibinfo {year} {2019})}\BibitemShut {NoStop}%
\bibitem [{\citenamefont {Wiersig}(2014)}]{wiersig2014enhancing}%
  \BibitemOpen
  \bibfield  {author} {\bibinfo {author} {\bibfnamefont {J.}~\bibnamefont
  {Wiersig}},\ }\bibfield  {title} {Enhancing the sensitivity of frequency and
  energy splitting detection by using exceptional points: application to
  microcavity sensors for single-particle detection,\ }\href@noop {} {\bibfield
   {journal} {\bibinfo  {journal} {Phys. Rev. Lett.}\ }\textbf {\bibinfo
  {volume} {112}},\ \bibinfo {pages} {203901} (\bibinfo {year}
  {2014})}\BibitemShut {NoStop}%
\bibitem [{\citenamefont {Hodaei}\ \emph {et~al.}(2017)\citenamefont {Hodaei},
  \citenamefont {Hassan}, \citenamefont {Wittek}, \citenamefont
  {Garcia-Gracia}, \citenamefont {El-Ganainy}, \citenamefont
  {Christodoulides},\ and\ \citenamefont {Khajavikhan}}]{hodaei2017enhanced}%
  \BibitemOpen
  \bibfield  {author} {\bibinfo {author} {\bibfnamefont {H.}~\bibnamefont
  {Hodaei}}, \bibinfo {author} {\bibfnamefont {A.~U.}\ \bibnamefont {Hassan}},
  \bibinfo {author} {\bibfnamefont {S.}~\bibnamefont {Wittek}}, \bibinfo
  {author} {\bibfnamefont {H.}~\bibnamefont {Garcia-Gracia}}, \bibinfo {author}
  {\bibfnamefont {R.}~\bibnamefont {El-Ganainy}}, \bibinfo {author}
  {\bibfnamefont {D.~N.}\ \bibnamefont {Christodoulides}}, \ and\ \bibinfo
  {author} {\bibfnamefont {M.}~\bibnamefont {Khajavikhan}},\ }\bibfield
  {title} {Enhanced sensitivity at higher-order exceptional points,\
  }\href@noop {} {\bibfield  {journal} {\bibinfo  {journal} {Nature}\ }\textbf
  {\bibinfo {volume} {548}},\ \bibinfo {pages} {187} (\bibinfo {year}
  {2017})}\BibitemShut {NoStop}%
\bibitem [{\citenamefont {Zhang}\ \emph {et~al.}(2019)\citenamefont {Zhang},
  \citenamefont {Ding}, \citenamefont {Zhou}, \citenamefont {Xu},\ and\
  \citenamefont {Jin}}]{zhang2019experimental}%
  \BibitemOpen
  \bibfield  {author} {\bibinfo {author} {\bibfnamefont {X.}~\bibnamefont
  {Zhang}}, \bibinfo {author} {\bibfnamefont {K.}~\bibnamefont {Ding}},
  \bibinfo {author} {\bibfnamefont {X.}~\bibnamefont {Zhou}}, \bibinfo {author}
  {\bibfnamefont {J.}~\bibnamefont {Xu}}, \ and\ \bibinfo {author}
  {\bibfnamefont {D.}~\bibnamefont {Jin}},\ }\bibfield  {title} {Experimental
  observation of an exceptional surface in synthetic dimensions with magnon
  polaritons,\ }\href@noop {} {\bibfield  {journal} {\bibinfo  {journal} {Phys.
  Rev. Lett.}\ }\textbf {\bibinfo {volume} {123}},\ \bibinfo {pages} {237202}
  (\bibinfo {year} {2019})}\BibitemShut {NoStop}%
\bibitem [{\citenamefont {Peng}\ \emph {et~al.}(2014)\citenamefont {Peng},
  \citenamefont {{\"O}zdemir}, \citenamefont {Lei}, \citenamefont {Monifi},
  \citenamefont {Gianfreda}, \citenamefont {Long}, \citenamefont {Fan},
  \citenamefont {Nori}, \citenamefont {Bender},\ and\ \citenamefont
  {Yang}}]{peng2014parity}%
  \BibitemOpen
  \bibfield  {author} {\bibinfo {author} {\bibfnamefont {B.}~\bibnamefont
  {Peng}}, \bibinfo {author} {\bibfnamefont {{\c{S}}.~K.}\ \bibnamefont
  {{\"O}zdemir}}, \bibinfo {author} {\bibfnamefont {F.}~\bibnamefont {Lei}},
  \bibinfo {author} {\bibfnamefont {F.}~\bibnamefont {Monifi}}, \bibinfo
  {author} {\bibfnamefont {M.}~\bibnamefont {Gianfreda}}, \bibinfo {author}
  {\bibfnamefont {G.~L.}\ \bibnamefont {Long}}, \bibinfo {author}
  {\bibfnamefont {S.}~\bibnamefont {Fan}}, \bibinfo {author} {\bibfnamefont
  {F.}~\bibnamefont {Nori}}, \bibinfo {author} {\bibfnamefont {C.~M.}\
  \bibnamefont {Bender}}, \ and\ \bibinfo {author} {\bibfnamefont
  {L.}~\bibnamefont {Yang}},\ }\bibfield  {title} {Parity--time-symmetric
  whispering-gallery microcavities,\ }\href@noop {} {\bibfield  {journal}
  {\bibinfo  {journal} {Nature Physics}\ }\textbf {\bibinfo {volume} {10}},\
  \bibinfo {pages} {394} (\bibinfo {year} {2014})}\BibitemShut {NoStop}%
\bibitem [{\citenamefont {Chen}\ \emph {et~al.}(2017)\citenamefont {Chen},
  \citenamefont {Kaya~{\"O}zdemir}, \citenamefont {Zhao}, \citenamefont
  {Wiersig},\ and\ \citenamefont {Yang}}]{chen2017exceptional}%
  \BibitemOpen
  \bibfield  {author} {\bibinfo {author} {\bibfnamefont {W.}~\bibnamefont
  {Chen}}, \bibinfo {author} {\bibfnamefont {{\c{S}}.}~\bibnamefont
  {Kaya~{\"O}zdemir}}, \bibinfo {author} {\bibfnamefont {G.}~\bibnamefont
  {Zhao}}, \bibinfo {author} {\bibfnamefont {J.}~\bibnamefont {Wiersig}}, \
  and\ \bibinfo {author} {\bibfnamefont {L.}~\bibnamefont {Yang}},\ }\bibfield
  {title} {Exceptional points enhance sensing in an optical microcavity,\
  }\href@noop {} {\bibfield  {journal} {\bibinfo  {journal} {Nature}\ }\textbf
  {\bibinfo {volume} {548}},\ \bibinfo {pages} {192} (\bibinfo {year}
  {2017})}\BibitemShut {NoStop}%
\bibitem [{\citenamefont {Pick}\ \emph {et~al.}(2017)\citenamefont {Pick},
  \citenamefont {Lin}, \citenamefont {Jin},\ and\ \citenamefont
  {Rodriguez}}]{pick2017enhanced}%
  \BibitemOpen
  \bibfield  {author} {\bibinfo {author} {\bibfnamefont {A.}~\bibnamefont
  {Pick}}, \bibinfo {author} {\bibfnamefont {Z.}~\bibnamefont {Lin}}, \bibinfo
  {author} {\bibfnamefont {W.}~\bibnamefont {Jin}}, \ and\ \bibinfo {author}
  {\bibfnamefont {A.~W.}\ \bibnamefont {Rodriguez}},\ }\bibfield  {title}
  {Enhanced nonlinear frequency conversion and purcell enhancement at
  exceptional points,\ }\href@noop {} {\bibfield  {journal} {\bibinfo
  {journal} {Physical Review B}\ }\textbf {\bibinfo {volume} {96}},\ \bibinfo
  {pages} {224303} (\bibinfo {year} {2017})}\BibitemShut {NoStop}%
\bibitem [{\citenamefont {Lin}\ \emph {et~al.}(2016)\citenamefont {Lin},
  \citenamefont {Pick}, \citenamefont {Lon\ifmmode~\check{c}\else
  \v{c}\fi{}ar},\ and\ \citenamefont {Rodriguez}}]{PhysRevLett.117.107402}%
  \BibitemOpen
  \bibfield  {author} {\bibinfo {author} {\bibfnamefont {Z.}~\bibnamefont
  {Lin}}, \bibinfo {author} {\bibfnamefont {A.}~\bibnamefont {Pick}}, \bibinfo
  {author} {\bibfnamefont {M.}~\bibnamefont {Lon\ifmmode~\check{c}\else
  \v{c}\fi{}ar}}, \ and\ \bibinfo {author} {\bibfnamefont {A.~W.}\ \bibnamefont
  {Rodriguez}},\ }\bibfield  {title} {Enhanced spontaneous emission at
  third-order dirac exceptional points in inverse-designed photonic crystals,\
  }\href@noop {} {\bibfield  {journal} {\bibinfo  {journal} {Phys. Rev. Lett.}\
  }\textbf {\bibinfo {volume} {117}},\ \bibinfo {pages} {107402} (\bibinfo
  {year} {2016})}\BibitemShut {NoStop}%
\bibitem [{\citenamefont {L\"u}\ \emph {et~al.}(2015)\citenamefont {L\"u},
  \citenamefont {Jing}, \citenamefont {Ma},\ and\ \citenamefont
  {Wu}}]{PhysRevLett.114.253601}%
  \BibitemOpen
  \bibfield  {author} {\bibinfo {author} {\bibfnamefont {X.-Y.}\ \bibnamefont
  {L\"u}}, \bibinfo {author} {\bibfnamefont {H.}~\bibnamefont {Jing}}, \bibinfo
  {author} {\bibfnamefont {J.-Y.}\ \bibnamefont {Ma}}, \ and\ \bibinfo {author}
  {\bibfnamefont {Y.}~\bibnamefont {Wu}},\ }\bibfield  {title}
  {$\mathcal{P}\mathcal{T}$-symmetry-breaking chaos in optomechanics,\
  }\href@noop {} {\bibfield  {journal} {\bibinfo  {journal} {Phys. Rev. Lett.}\
  }\textbf {\bibinfo {volume} {114}},\ \bibinfo {pages} {253601} (\bibinfo
  {year} {2015})}\BibitemShut {NoStop}%
\bibitem [{\citenamefont {Bu}\ \emph {et~al.}(2023)\citenamefont {Bu},
  \citenamefont {Zhang}, \citenamefont {Ding}, \citenamefont {Li},
  \citenamefont {Zhang}, \citenamefont {Wang}, \citenamefont {Ding},
  \citenamefont {Yuan}, \citenamefont {Chen}, \citenamefont {\"Ozdemir},
  \citenamefont {Zhou}, \citenamefont {Jing},\ and\ \citenamefont
  {Feng}}]{PhysRevLett.130.110402}%
  \BibitemOpen
  \bibfield  {author} {\bibinfo {author} {\bibfnamefont {J.-T.}\ \bibnamefont
  {Bu}}, \bibinfo {author} {\bibfnamefont {J.-Q.}\ \bibnamefont {Zhang}},
  \bibinfo {author} {\bibfnamefont {G.-Y.}\ \bibnamefont {Ding}}, \bibinfo
  {author} {\bibfnamefont {J.-C.}\ \bibnamefont {Li}}, \bibinfo {author}
  {\bibfnamefont {J.-W.}\ \bibnamefont {Zhang}}, \bibinfo {author}
  {\bibfnamefont {B.}~\bibnamefont {Wang}}, \bibinfo {author} {\bibfnamefont
  {W.-Q.}\ \bibnamefont {Ding}}, \bibinfo {author} {\bibfnamefont {W.-F.}\
  \bibnamefont {Yuan}}, \bibinfo {author} {\bibfnamefont {L.}~\bibnamefont
  {Chen}}, \bibinfo {author} {\bibfnamefont {i.~m. c.~K.}\ \bibnamefont
  {\"Ozdemir}}, \bibinfo {author} {\bibfnamefont {F.}~\bibnamefont {Zhou}},
  \bibinfo {author} {\bibfnamefont {H.}~\bibnamefont {Jing}}, \ and\ \bibinfo
  {author} {\bibfnamefont {M.}~\bibnamefont {Feng}},\ }\bibfield  {title}
  {Enhancement of quantum heat engine by encircling a liouvillian exceptional
  point,\ }\href@noop {} {\bibfield  {journal} {\bibinfo  {journal} {Phys. Rev.
  Lett.}\ }\textbf {\bibinfo {volume} {130}},\ \bibinfo {pages} {110402}
  (\bibinfo {year} {2023})}\BibitemShut {NoStop}%
\bibitem [{\citenamefont {Wang}(2022)}]{SM}%
  \BibitemOpen
  \bibfield  {author} {\bibinfo {author} {\bibfnamefont {C.}~\bibnamefont
  {Wang}},\ }\bibfield  {title} {Supplementary material for "giant enhancement
  of magnonic frequency combs by exceptional points",\ }\href@noop {} {\
  (\bibinfo {year} {2022})}\BibitemShut {NoStop}%
\bibitem [{\citenamefont {Sasaki}\ \emph {et~al.}(2021)\citenamefont {Sasaki},
  \citenamefont {Nii},\ and\ \citenamefont {Onose}}]{sasaki2021magnetization}%
  \BibitemOpen
  \bibfield  {author} {\bibinfo {author} {\bibfnamefont {R.}~\bibnamefont
  {Sasaki}}, \bibinfo {author} {\bibfnamefont {Y.}~\bibnamefont {Nii}}, \ and\
  \bibinfo {author} {\bibfnamefont {Y.}~\bibnamefont {Onose}},\ }\bibfield
  {title} {Magnetization control by angular momentum transfer from surface
  acoustic wave to ferromagnetic spin moments,\ }\href@noop {} {\bibfield
  {journal} {\bibinfo  {journal} {Nature Communications}\ }\textbf {\bibinfo
  {volume} {12}},\ \bibinfo {pages} {2599} (\bibinfo {year}
  {2021})}\BibitemShut {NoStop}%
\bibitem [{\citenamefont {Lodahl}\ \emph {et~al.}(2017)\citenamefont {Lodahl},
  \citenamefont {Mahmoodian}, \citenamefont {Stobbe}, \citenamefont
  {Rauschenbeutel}, \citenamefont {Schneeweiss}, \citenamefont {Volz},
  \citenamefont {Pichler},\ and\ \citenamefont {Zoller}}]{lodahl2017chiral}%
  \BibitemOpen
  \bibfield  {author} {\bibinfo {author} {\bibfnamefont {P.}~\bibnamefont
  {Lodahl}}, \bibinfo {author} {\bibfnamefont {S.}~\bibnamefont {Mahmoodian}},
  \bibinfo {author} {\bibfnamefont {S.}~\bibnamefont {Stobbe}}, \bibinfo
  {author} {\bibfnamefont {A.}~\bibnamefont {Rauschenbeutel}}, \bibinfo
  {author} {\bibfnamefont {P.}~\bibnamefont {Schneeweiss}}, \bibinfo {author}
  {\bibfnamefont {J.}~\bibnamefont {Volz}}, \bibinfo {author} {\bibfnamefont
  {H.}~\bibnamefont {Pichler}}, \ and\ \bibinfo {author} {\bibfnamefont
  {P.}~\bibnamefont {Zoller}},\ }\bibfield  {title} {Chiral quantum optics,\
  }\href@noop {} {\bibfield  {journal} {\bibinfo  {journal} {Nature}\ }\textbf
  {\bibinfo {volume} {541}},\ \bibinfo {pages} {473} (\bibinfo {year}
  {2017})}\BibitemShut {NoStop}%
\bibitem [{\citenamefont {Yu}\ \emph {et~al.}(2023)\citenamefont {Yu},
  \citenamefont {Luo},\ and\ \citenamefont {Bauer}}]{yu2023chirality}%
  \BibitemOpen
  \bibfield  {author} {\bibinfo {author} {\bibfnamefont {T.}~\bibnamefont
  {Yu}}, \bibinfo {author} {\bibfnamefont {Z.}~\bibnamefont {Luo}}, \ and\
  \bibinfo {author} {\bibfnamefont {G.~E.}\ \bibnamefont {Bauer}},\ }\bibfield
  {title} {Chirality as generalized spin--orbit interaction in spintronics,\
  }\href@noop {} {\bibfield  {journal} {\bibinfo  {journal} {Physics Reports}\
  }\textbf {\bibinfo {volume} {1009}},\ \bibinfo {pages} {1} (\bibinfo {year}
  {2023})}\BibitemShut {NoStop}%
\bibitem [{\citenamefont {Yu}\ \emph {et~al.}(2020)\citenamefont {Yu},
  \citenamefont {Zhang}, \citenamefont {Sharma}, \citenamefont {Zhang},
  \citenamefont {Blanter},\ and\ \citenamefont {Bauer}}]{yu2020magnon}%
  \BibitemOpen
  \bibfield  {author} {\bibinfo {author} {\bibfnamefont {T.}~\bibnamefont
  {Yu}}, \bibinfo {author} {\bibfnamefont {Y.-X.}\ \bibnamefont {Zhang}},
  \bibinfo {author} {\bibfnamefont {S.}~\bibnamefont {Sharma}}, \bibinfo
  {author} {\bibfnamefont {X.}~\bibnamefont {Zhang}}, \bibinfo {author}
  {\bibfnamefont {Y.~M.}\ \bibnamefont {Blanter}}, \ and\ \bibinfo {author}
  {\bibfnamefont {G.~E.}\ \bibnamefont {Bauer}},\ }\bibfield  {title} {Magnon
  accumulation in chirally coupled magnets,\ }\href@noop {} {\bibfield
  {journal} {\bibinfo  {journal} {Phys. Rev. Lett.}\ }\textbf {\bibinfo
  {volume} {124}},\ \bibinfo {pages} {107202} (\bibinfo {year}
  {2020})}\BibitemShut {NoStop}%
\bibitem [{\citenamefont {Else}\ \emph {et~al.}(2016)\citenamefont {Else},
  \citenamefont {Bauer},\ and\ \citenamefont {Nayak}}]{else2016floquet}%
  \BibitemOpen
  \bibfield  {author} {\bibinfo {author} {\bibfnamefont {D.~V.}\ \bibnamefont
  {Else}}, \bibinfo {author} {\bibfnamefont {B.}~\bibnamefont {Bauer}}, \ and\
  \bibinfo {author} {\bibfnamefont {C.}~\bibnamefont {Nayak}},\ }\bibfield
  {title} {Floquet time crystals,\ }\href@noop {} {\bibfield  {journal}
  {\bibinfo  {journal} {Phys. Rev. Lett.}\ }\textbf {\bibinfo {volume} {117}},\
  \bibinfo {pages} {090402} (\bibinfo {year} {2016})}\BibitemShut {NoStop}%
\end{thebibliography}
\end{document}